\documentclass[a4paper,paper]{JHEP3}

\usepackage{epsfig,multicol}

\title{Extremal black holes in $D=5$: SUSY vs. Gauss-Bonnet
 corrections}

\author{Maro Cvitan$^a$, Predrag Dominis Prester$^{a,b}$,
  Silvio Pallua$^a$ and Ivica Smoli\'{c}$^a$\\
 $^a$Theoretical Physics Department, Faculty of Science, University of
  Zagreb\\ p.p. 331, HR-10002 Zagreb, Croatia\\
 $^b$ Physics Department, Faculty of Arts and Sciences,
  University of Rijeka\\ Omladinska 14, HR-51000 Rijeka, Croatia\\ 
 E-mail: \email{mcvitan@phy.hr}, \email{pprester@phy.hr},
  \email{ismolic@phy.hr}, \email{pallua@phy.hr}}

\preprint{arXiv:0706.1167 [hep-th]
          }

\abstract{We analyse near-horizon solutions and compare the results
for the black hole entropy of five-dimensional spherically symmetric 
extremal black holes
when the $N=2$ SUGRA actions are supplied with two different types of
higher-order corrections: (1) supersymmetric completion of gravitational
Chern-Simons term, and (2) Gauss-Bonnet term. We show that for large BPS 
black holes lowest order
$\alpha'$ corrections to the entropy are the same, but for non-BPS are
generally different. We pay special attention to the class of
prepotentials connected with $K3\times T^2$ and $T^6$
compactifications. For supersymmetric correction we find beside BPS
also a set of non-BPS solutions. In the particular case of $T^6$
compactification (equivalent to the heterotic string on $T^4\times S^1$)
we find the (almost) complete set of solutions (with exception of some
non-BPS small black holes), and show that entropy of small black holes
is different from statistical entropy obtained by counting of
microstates of heterotic string theory. We also find complete set of
solutions for $K3\times T^2$ and $T^6$ case when correction is given
by Gauss-Bonnet term. Contrary to four-dimensional case, obtained
entropy is different from the one with supersymmetric correction. We
show that in Gauss-Bonnet case entropy of small ``BPS''  black
holes agrees with microscopic entropy in the known cases.}

\begin{document}

\section{Introduction}
\label{sec:intro}

In recent years a lot of attention was directed towards higher
curvature corrections in effective SUGRA field theories appearing in
compactifications of string theories. Particularly interesting
question is how these corrections are affecting black hole solutions,
and in particular their entropies. One of the main successes so far of 
string theory is that it offers statistical explanation of black
hole entropy by direct counting of microstates. In some cases it
was possible to obtain not only lowest order Bekenstein-Hawking area
law, but also higher corrections in string tension $\alpha'$, and even
$\alpha'$ exact expressions for the entropy. These calculations are 
typically performed in the limit of small string coupling constant 
$g_s$ in the realm of perturbative string theory, where space-time is
almost flat and black holes are actually not present. It is expected
that these objects become black holes when one turns on $g_s$ enough
so that their size becomes smaller than their corresponding
Schwarzschild radius. Unfortunately, it is not known how to make
direct calculations in string theory in this regime. However, when one
goes in the opposite extreme where the Schwarzschild radius becomes
much larger than the string length $\ell_s = \sqrt{\alpha'}$,
then one can use low energy effective action where black holes appear
as classical solutions. 

The situation is especially interesting for
BPS black holes. In this case on the perturbative string side one is
counting number of states in short multiplets, which is expected to
not depend on $g_s$, at least generically (this property can be
violated in special circumstances like, e.g., short multiplet 
crossings). This means that one can directly compare statistical (or
microscopic) entropy from perturbative string and macroscopic entropy
from classical supergravity. By comparing the results from the both
limits we have not only succeeded to do sophisticated perturbative
consistency checks on the theory, but also improved our understanding
both of string theory and supergravity. Developments include attractor
mechanism and relation to topological strings \cite{topstr}. Especially
fruitful and rich are results obtained for black holes in $D=4$ (for 
reviews see \cite{D4bh}).

In $D=4$ especially nice examples are provided by heterotic 
string compactified on $K3\times S^1\times S^1$ or 
$T^4\times S^1\times S^1$ \cite{Sen:2005iz}. The simplest BPS states
correspond to large spherically symmetric black holes having 4 charges 
(2 electric and 2 magnetic), for which statistical entropy was found 
\cite{LopesCardoso:2004xf,Jatkar:2005bh,David:2006ji,David:2006yn}.  
The macroscopic black hole entropy was calculated using two types of
actions with higher order $R^2$ terms -- supersymmetric and 
Gauss-Bonnet. In the regime where $g_s$ is small near the horizon 
(limit where electric charges are much larger than magnetic) all 
results are {\em exactly} equal (i.e., in all orders in $\alpha'$). This 
is surprising because in both of these effective actions one has 
neglected an infinite number of terms in low energy effective action and
one would at best expect agreement in first order in $\alpha'$. There is an
argumentation \cite{KraLar,David:2007ak}, based on $AdS_3$ arguments, 
which explains why 
corrections of higher order than $R^2$ are irrelevant for calculation of 
black hole entropy, but it still does not explain why these two particular 
types of corrections are working for BPS black holes. 

These matches are even more surprising when one takes magnetic charges to
be zero. One gets 2-charge {\em small} black holes which in the lowest order
have null-singular horizon with vanishing area, which is made regular by
inclusion of higher curvature corrections \cite{smallBH,Sen:2004}. As curvature
is of order $1/\alpha'$, all terms in the effective action give a priori 
contribution to the entropy which is of the same order in $\alpha'$. This is a
consequence of the fact that here we are naively outside of the regime where 
effective action should be applicable.

In view of these results, it would be interesting to consider what happens in
higher dimensions $D>4$. 2-charge BPS states and corresponding small extremal 
black holes generalize to all $D\leq9$. In \cite{Sen:2005kj} it was shown that
simple Gauss-Bonnet correction gives correct result for the entropy of such 
black holes also in $D=5$, but not for $D>5$. Afterwards, in 
\cite{Prester:2005qs} it was shown that there is an effective action where 
higher order terms are given by linear combination of all generalized 
Gauss-Bonnet densities (with uniquely fixed coefficients) which gives the 
correct entropy for all dimensions. For large black holes things do not 
generalize directly. In $D=5$ simplest are 3-charge BPS black holes, but even 
for them statistical entropy is known only in lowest order in $\alpha'$ 
\cite{Strominger:1996}. Let us
mention that the argumentation based on $AdS_3$ geometry has not been 
generalized to $D>4$.

Motivated by all this, in this paper we analyse near-horizon solutions and 
calculate macroscopic entropy for a class of five-dimensional black holes in 
the $N=2$ supergravities for which higher-derivative $R^2$ actions were 
recently obtained in \cite{Hanaki:2006pj}. In Sec.\ 2 we present 
$D=5$ supersymmetric action \cite{Hanaki:2006pj}. In Sec.\ 3 we review Sen's 
entropy function formalism \cite{Sen:2005wa}. In Sec.\ 4 we present maximally
supersymmetric $AdS_2\times S^3$ solution which describes near-horizon geometry
of purely electrically charged 1/2 BPS black holes. In Sec.\ 5 for the case of
simple $STU$ prepotential we find non-BPS solutions for all values of charges, 
except for some small black holes with one charge equal to 0 or $\pm1$. In 
Sec.\ 6 we show how and when solutions from Sec.\ 5 can be generalized. In 
Sec.\ 7 we present near horizon solutions for 3-charge black holes in 
heterotic string theory compactified on $K3\times S^1$ when the $R^2$ 
correction is given by Gauss-Bonnet density. and compare them with the 
results from SUSY action. We show that for small black holes Gauss-Bonnet 
correction keeps producing results in agreement with microscopic analyses.
In Appendix A we present generalisation of Sec.\ 5 to general correction
coefficients $c_I$, and in Appendix B derivations of results presented in 
Sec.\ 7.

While our work was in the late stages references
\cite{Castro:2007hc,Alishahiha:2007nn}
appeared which have some overlap with our paper. In these papers 
near-horizon solutions and the entropy for BPS black holes for
supersymmetric corrections were given, which is a subject of our 
Sec.\ \ref{sec:bps}. Our results are in agreement with those in
\cite{Castro:2007hc,Alishahiha:2007nn}. However, we emphasize that 
our near-horizon solutions in Secs.\ \ref{sec:stu} and \ref{sec:genpp}
for non-BPS black holes are completely new. Also, in 
\cite{Alishahiha:2007nn} there is a statement on matching of the 
entropy of BPS black hole for supersymmetric and Gauss-Bonnet 
correction. We explicitly show in Sec.\ \ref{sec:GB} that this is 
valid just for first $\alpha'$ correction.

\section{Higher derivative $N=2$ SUGRA in $D=5$}
\label{sec:sugra}

Bosonic part of the Lagrangian for the $N=2$ supergravity action in 
five dimensions is given by
\begin{eqnarray} \label{l0susy}
4\pi^2\mathcal{L}_0 &=& 2 \partial^a \mathcal{A}^\alpha_i \partial_a
\mathcal{A}_\alpha^i + \mathcal{A}^2 
\left(\frac{D}{4}-\frac{3}{8}R-\frac{v^2}{2}\right)
+ \mathcal{N} \left(\frac{D}{2}+\frac{R}{4}+3v^2\right)
+ 2 \mathcal{N}_I v^{ab} F_{ab}^I \nonumber \\ 
&& + \mathcal{N}_{IJ} \left( \frac{1}{4} F_{ab}^I F^{Jab} 
 + \frac{1}{2} \partial_a M^I \partial^a M^J \right)
+ \frac{e^{-1}}{24} c_{IJK} A_a^I F_{bc}^J F_{de}^K \epsilon^{abcde}
\end{eqnarray}
where $\mathcal{A}^2 = \mathcal{A}^\alpha_{i\,ab}
\mathcal{A}_\alpha^{i\,ab}$ and $v^2 = v_{ab}v^{ab}$. Also,
\begin{equation}
\mathcal{N} = \frac{1}{6} c_{IJK} M^I M^J M^K , \quad
\mathcal{N}_I = \partial_I \mathcal{N} = \frac{1}{2} c_{IJK} M^J M^K
, \quad 
\mathcal{N}_{IJ} = \partial_I \partial_J \mathcal{N} = c_{IJK} M^K
\end{equation}

A bosonic field content of the theory is the following. We have Weyl
multiplet which contains the f\"{u}nfbein $e_\mu^a$, the two-form 
auxiliary field $v_{ab}$, and the scalar auxiliary field $D$. There
are $n_V$ vector multiplets enumerated by $I=1,\ldots,n_V$, each
containing the one-form gauge field $A^I$ (with the two-form field 
strength $F^I=dA^I$) and the scalar $M^I$. Scalar fields 
$\mathcal{A}_\alpha^i$, which are belonging to the hypermultiplet, can
be gauge fixed and the convenient choice is given by
\begin{equation} \label{hgfix}
\mathcal{A}^2 = -2 \;, \qquad \partial_a \mathcal{A}^\alpha_i = 0
\end{equation}

One can use equations of motion for auxiliary fields to get rid of
them completely and obtain the Lagrangian in a standard form:
\begin{equation} \label{l0noaux}
4\pi^2\mathcal{L}_0 = R - G_{IJ} \partial_a M^I \partial^a M^J
- \frac{1}{2} G_{IJ} F_{ab}^I F^{Jab}
+ \frac{e^{-1}}{24} c_{IJK} A_a^I F_{bc}^J F_{de}^K \epsilon^{abcde}
\end{equation}
with
\begin{equation} \label{gIJ}
G_{IJ} = - \frac{1}{2} \partial_I\partial_J(\ln \mathcal{N}) = 
\frac{1}{2} \left( \mathcal{N}_I \mathcal{N}_J - \mathcal{N}_{IJ}
\right)
\end{equation}
and where $\mathcal{N}=1$ is implicitly understood (but only after taking 
derivatives in (\ref{gIJ})). We shall later use this form of Lagrangian
to make connection with heterotic string effective actions.

Lagrangian (\ref{l0noaux}) can be obtained from 11-dimensional SUGRA by
compactifying on six-dimensional Calabi-Yau spaces ($CY_3$). Then $M^I$
have interpretation as moduli (volumes of $(1,1)$-cycles), and $c_{IJK}$
as intersection numbers. Condition $\mathcal{N}=1$ is a condition of real
special geometry. For a recent review and further references see 
\cite{Larsen:2006}. 

Action (\ref{l0susy}) is invariant under SUSY variations, which
when acting on the purely bosonic configurations (and after using 
(\ref{hgfix})) are given with
\begin{eqnarray} \label{svar}
\delta\psi_\mu^i &=& \mathcal{D}_\mu\varepsilon^i + \frac{1}{2}v^{ab}
 \gamma_{\mu ab}\varepsilon^i - \gamma_\mu\eta^i \nonumber \\
\delta\xi^i &=& D\varepsilon^i 
 - 2\gamma^c\gamma^{ab}\varepsilon^i\mathcal{D}_a v_{bc}
 - 2\gamma^a\varepsilon^i\epsilon_{abcde}v^{bc}v^{de} 
 + 4\gamma\cdot v\eta^i \nonumber \\
\delta\Omega^{Ii} &=& - \frac{1}{4}\gamma\cdot F^{I}\varepsilon^i
 - \frac{1}{2}\gamma^a\partial_a M^{I}\varepsilon^i - M^{I}\eta^i
 \nonumber \\
\delta\zeta^{\alpha} &=& 
 \left(3\eta^j-\gamma\cdot v\varepsilon^j\right)\mathcal{A}_j^\alpha
\end{eqnarray}
where $\psi_\mu^i$ is gravitino, $\xi^i$ auxiliary Majorana spinor
(Weyl multiplet), $\delta\Omega^{Ii}$ gaugino (vector multiplets), and
$\zeta^{\alpha}$ is a fermion field from hypermultiplet.

In \cite{Hanaki:2006pj} four derivative part of the action was
constructed by supersymmetric completion of the mixed
gauge-gravitational Chern-Simons term 
$A \land \textrm{tr} (R \land R)$. The bosonic part of the action 
relevant for our purposes was shown to be
\begin{eqnarray} \label{l1susy}
4\pi^2\mathcal{L}_1 &=& \frac{c_{I}}{24} \left\{ \frac{e^{-1}}{16}
\epsilon_{abcde} A^{Ia} C^{bcfg} C^{de}_{\;\;\;\,fg} 
+ M^I \left[ \frac{1}{8} C^{abcd} C_{abcd} + \frac{1}{12} D^2 
 - \frac{1}{3} C_{abcd} v^{ab} v^{cd} 
\right. \right. \nonumber \\ &&
 + 4 v_{ab}v^{bc} v_{cd} v^{da} - (v_{ab}v^{ab})^2
 + \frac{8}{3} v_{ab} \hat{\mathcal{D}}^b \hat{\mathcal{D}}_c v^{ac}
 + \frac{4}{3} \hat{\mathcal{D}}^a v^{bc} \hat{\mathcal{D}}_a v_{bc}
 + \frac{4}{3} \hat{\mathcal{D}}^a v^{bc} \hat{\mathcal{D}}_b v_{ca}
\nonumber \\ && \left. 
 - \frac{2}{3} e^{-1} \epsilon_{abcde} v^{ab} v^{cd}
   \hat{\mathcal{D}}_f v^{ef} \right] 
+ F^{Iab} \left[ \frac{1}{6} v_{ab} D - \frac{1}{2} C_{abcd} v^{cd}
 + \frac{2}{3} e^{-1} \epsilon_{abcde} v^{cd} 
   \hat{\mathcal{D}}_f v^{ef} 
\right. \nonumber \\ && \left. \left.
 + e^{-1} \epsilon_{abcde} v^{c}_{\;f} \hat{\mathcal{D}}^d v^{ef}
 - \frac{4}{3} v_{ac}v^{cd} v_{db} - \frac{1}{3} v_{ab} v^2 \right]
\right\}
\end{eqnarray}
where $c_I$ are some constant coefficients\footnote{From the viewpoint of
compactification of $D=11$ SUGRA they are topological numbers connected
to second Chern class, see \cite{Ferrara:1996hh}.}, 
$C_{abcd}$ is the Weyl tensor which in five dimensions is
\begin{equation}
C^{ab}_{\;\;\;\,cd} = R^{ab}_{\;\;\;\,cd} - \frac{1}{3}
\left( g^a_c R^b_d - g^a_d R^b_c - g^b_c R^a_d + g^b_d R^a_c \right)
+ \frac{1}{12} R \left( g^a_c g^b_d - g^a_d g^b_c \right)
\end{equation}
and $\hat{\mathcal{D}}_a$ is the conformal covariant derivative, which
when appearing linearly in (\ref{l1susy}) can be substituted with 
ordinary covariant derivative $\mathcal{D}_a$, but when taken twice 
produces additional curvature contributions \cite{Castro:2007sd}:
\begin{equation}
v_{ab} \hat{\mathcal{D}}^b \hat{\mathcal{D}}_c v^{ac} = 
v_{ab} \mathcal{D}^b \mathcal{D}_c v^{ac} 
+ \frac{2}{3} v^{ac} v_{cb} R_a^b + \frac{1}{12} v^2 R
\end{equation}

We are going to analyse extremal black hole solutions of the action
obtained by combining (\ref{l0susy}) and (\ref{l1susy}):\footnote{Our
conventions for Newton coupling is $G_5=\pi^2/4$ and for the string
tension $\alpha'=1$.}
\begin{equation} \label{lsusy}
\mathcal{A} = \int dx^5 \sqrt{-g} \mathcal{L} 
 = \int dx^5 \sqrt{-g} (\mathcal{L}_0 + \mathcal{L}_1)
\end{equation}
As (\ref{l1susy}) is a complicated function of auxiliary fields
(including derivatives) it is now impossible to integrate them out in
the closed form and obtain an action which includes just the physical
fields. 

\section{Near horizon geometry and entropy function formalism}
\label{sec:ent-func}

The action (\ref{lsusy}) is quartic in derivatives and generally
probably too complicated for finding complete analytical black hole
solutions even in the simplest spherically symmetric case. But, if one
is more modest and interested just in a near-horizon behavior (which is
enough to find the entropy) of {\em extremal} black holes, there is a
smart way to do the job - Sen's entropy function formalism
\cite{Sen:2005wa}.\footnote{This formalism was used recently in 
near-horizon analyses of a broad classes of black holes and higher 
dimensional objects \cite{entfappl}. For generalisation to rotating black 
holes see \cite{Astefanesei:2006dd}. For comparison with SUSY entropy functions
see \cite{Cardoso:2006xz}.}

For five-dimensional spherically symmetric extremal black holes 
near-horizon geometry is expected to be $AdS_2\times S^3$, which has 
$SO(2,1)\times SO(4)$ symmetry \cite{Kunduri:2007vf}. If the Lagrangian 
can be written in a 
manifestly diffeomorphism covariant and gauge invariant way (as a 
function of metric, Riemann tensor, covariant derivative, and gauge 
invariant fields, but without connections) it is expected that near 
the horizon the complete background should respect this symmetry. Then
it follows that the only fields which can acquire non-vanishing values
near the horizon are scalars $\phi_s$, (purely electric) two-forms 
fields $F^{I}$, and (purely magnetic) three-form fields $H_{m}$. 
Explicitly written:
\begin{eqnarray} \label{efgen}
&& ds^2 = v_1 \left( -x^2 dt^2 + \frac{dx^2}{x^2} \right) 
 + v_2\,d\Omega_{3}^2 \nonumber \\
&& \phi_s = u_s \;, \qquad s=1,\ldots,n_s  \nonumber \\
&& F^{I} = - e^I \mathbf{\epsilon}_{A} \;, \qquad I=1,\ldots,n_F
 \nonumber \\
&& H_{m} = 2q_m \mathbf{\epsilon}_{S} \;, \qquad m=1,\ldots,n_H
\end{eqnarray}
where $v_{1,2}$, $u_s$, $e^I$ and $q_m$ are constants, 
$\mathbf{\epsilon}_{A}$ and $\mathbf{\epsilon}_{S}$ are induced
volume-forms on $AdS_2$ and $S^3$, respectively. In case where 
$F^{I}$ ($H_{m}$) are gauge field strengths, $e^I$ ($q_m$) are 
electric field strengths (magnetic charges).

It is important to notice that all covariant derivatives in this 
background are vanishing.

To obtain near-horizon solutions one defines
\begin{equation}\label{fdef}
f(\vec{v},\vec{u},\vec{e}) = \int_{S^3} \sqrt{-g} \, \mathcal{L}
\end{equation}
extremization of which over $\vec{v}$ and $\vec{u}$ gives equations of
motion (EOM's)
\begin{equation} \label{eomf}
\frac{\partial f}{\partial v_i}=0 \;, \qquad
\frac{\partial f}{\partial u_s}=0 \;,
\end{equation}
and derivatives over $\vec{e}$ are giving (properly normalized)
electric charges:
\begin{equation}\label{chgdef}
q_I = \frac{\partial f}{\partial e^I}
\end{equation}
Finally, the entropy (equal to the Wald formula \cite{Wald}) is given 
with
\begin{equation} \label{entropy}
S_{BH} = 2\pi \left( q_I \, e^I - f \right)
\end{equation}

Equivalently, one can define entropy function $F$ as a Legendre 
transform of the function $f$ with respect to the electric fields and
charges
\begin{equation}
F(\vec{v},\vec{u},\vec{e},\vec{q}) = 2\pi \left( q_I \, e^I 
 - f(\vec{v},\vec{u},\vec{e}) \right)
\end{equation}
Now equations of motion are obtained by extremizing entropy function
\begin{equation} \label{eomF}
0 = \frac{\partial F}{\partial v_i} \;, \qquad
0 = \frac{\partial F}{\partial u_s} \;, \qquad
0 = \frac{\partial F}{\partial e^I}
\end{equation}
and the value at the extremum gives the black hole entropy
\begin{equation}
S_{BH} = F(\vec{v},\vec{u},\vec{e},\vec{q}) \qquad \mbox{when }
 \vec{v},\vec{u},\vec{e} \mbox{ satisfy (\ref{eomF})}
\end{equation}

We want next to apply entropy function formalism to the $N=2$ SUGRA from
Sec.\ \ref{sec:sugra}. In this case for the near-horizon geometry 
(\ref{efgen}) we explicitly have
\begin{eqnarray} \label{efhere}
&& ds^2 = v_1 \left( -x^2 dt^2 + \frac{dx^2}{x^2} \right)
 + v_2\,d\Omega_{3}^2 \nonumber \\
&& F^{I}_{tr}(x) = -e^I \;,\qquad v_{tr}(x) = V \nonumber \\
&& M^I(x) = M^I \;, \qquad D(x) = D
\end{eqnarray}
where $v_i$, $e^I$, $M^I$, $V$, and $D$ are constants.

Putting (\ref{efhere}) into (\ref{l0susy}) and (\ref{l1susy}) one gets
\begin{eqnarray} \label{f0susy}
f_0 &=& 
\frac{1}{4} \sqrt{v_2}
\left[\left(\mathcal{N}+3\right) \left(3v_1-v_2\right)-
4 V^2 \left(3\mathcal{N}+1\right) \frac{v_2}{v_1} \right.
\nonumber \\ && \left. 
 \qquad\quad + 8 V \mathcal{N}_i e^i \frac{ v_2}{v_1}
- \mathcal{N}_{ij} e^i e^j \frac{v_2}{v_1}
+ D (\mathcal{N}-1)v_1 v_2 \right] 
\end{eqnarray}
and
\begin{eqnarray} \label{f1susy}
f_1 &=& v_1  v_2^{3/2}
\left\{
\frac{c_I e^I}{48} 
\left[-\frac{4 V^3}{3 v_1^4}+\frac{D V}{3 v_1^2}+
\frac{V}{v_1^2}\left(\frac{1}{v_1}-\frac{1}{v_2}\right)\right]
\phantom{\left(\frac{1}{v_1}\right)^2} \right.
\nonumber \\
&& + \left. 
\frac{c_I M^I}{48} 
\left[\frac{D^2}{12}+\frac{4 V^4}{v_1^4}+
\frac{1}{4} \left(\frac{1}{v_1}-\frac{1}{v_2}\right)^2-
\frac{2 V^2}{3v_1^2}\left(\frac{5}{v_1}+\frac{3}{v_2}\right)\right]
\right\} \;,
\end{eqnarray}
correspondingly. Notice that for the background (\ref{efhere}) all terms
containing $\varepsilon_{abcde}$ tensor vanish. Complete function $f$ is
a sum
\begin{equation} \label{fsusy}
f = f_0 + f_1
\end{equation}
and EOM's near the horizon are equivalent to
\begin{equation} \label{seom}
0 = \frac{\partial f}{\partial v_1} \;, \qquad
0 = \frac{\partial f}{\partial v_2} \;, \qquad
0 = \frac{\partial f}{\partial M^I} \;, \qquad
0 = \frac{\partial f}{\partial V} \;, \qquad
0 = \frac{\partial f}{\partial D} \;.
\end{equation}

Notice that both $f_0$ and $f_1$ (and so $f$) are invariant on the
transformation $e^I\to-e^I$, $V\to-V$, with other variables
remaining the same. This symmetry follows from CPT invariance. We
shall use it to obtain new solutions with $q_I\to-q_I$.

\section{Solutions with maximal supersymmetry}
\label{sec:bps}

We want to find near horizon solutions using entropy function
formalism described in Sec.\ \ref{sec:ent-func}. The procedure is to
fix the set of electric charges $q_I$ and then solve the system of
equations (\ref{seom}), (\ref{chgdef}) with the function $f$ given by
(\ref{f0susy}), (\ref{f1susy}), (\ref{fsusy}). It is immediately 
obvious that though the system is algebraic, it is in generic case too
complicated to be solved in direct manner, and that one should try to
find some additional information.

Such additional information can be obtained from supersymmetry. It is
known that there should be 1/2 BPS black hole solutions, for which it
was shown in \cite{Chamseddine:1996pi} that near the horizon
supersymmetry is enhanced fully. This means that in this case we can
put all variations 
in (\ref{svar}) to zero, which for $AdS_2\times S^3$ background become
\begin{eqnarray} \label{svarBH}
0 &=& \mathcal{D}_\mu\varepsilon^i + \frac{1}{2}v^{ab}
 \gamma_{\mu ab}\varepsilon^i - \gamma_\mu\eta^i \nonumber \\
0 &=& D\varepsilon^i + 4\,\gamma\cdot v\,\eta^i \nonumber \\
0 &=& - \frac{1}{4}\gamma\cdot F^{I}\varepsilon^i - M^{I}\eta^i
 \nonumber \\
0 &=& \left(3\eta^j-\gamma\cdot v\,\varepsilon^j\right)
 \mathcal{A}_j^\alpha
\end{eqnarray}
Last equation fixes the spinor parameter $\eta$ to be
\begin{equation} \label{etacond}
\eta^j = \frac{1}{3}(\gamma\cdot v)\varepsilon^j
\end{equation}
Using this, and the condition that $\varepsilon^i$ is (geometrical) 
Killing spinor, in the remaining equations one gets\footnote{As
the detailed derivation was already presented in 
\cite{Castro:2007hc} (solutions in the whole space) and in 
\cite{Alishahiha:2007nn} (near horizon solutions),
we shall just state the results here.} the following conditions
\begin{equation} \label{bpscond}
v_2 = 4v_1 \;, \qquad M^I = \frac{e^I}{\sqrt{v_1}} \;, \qquad 
D = -\frac{3}{v_1} \;, \qquad V = \frac{3}{4}\sqrt{v_1}
\end{equation}
We see that conditions for full supersymmetry are so constraining that
they fix everything except one unknown, which we took above to be
$v_1$. To fix it, we just need one equation from (\ref{seom}). In our
case the simplest is to take equation for $D$, which gives
\begin{equation} \label{v1bps}
v_1^{3/2} = (e)^3 - \frac{c_{I}e^I}{48}
\end{equation}
where we used a notation
\begin{equation}
(e)^3 \equiv \frac{1}{6} c_{IJK} e^I e^J e^K
\end{equation}
We note that higher derivative corrections violate real special geometry
condition, i.e., we have now $\mathcal{N}\neq1$.\footnote{We emphasize
that one should be cautious in geometric interpretation of this result. Higher
order corrections generally change relations between fields in the effective 
action and geometric moduli, and one needs field redefinitions to restore
the relations. Then correctly defined moduli may still satisfy condition 
for real special geometry.}

Using (\ref{bpscond}) and (\ref{v1bps}) in the expression for the
entropy (\ref{entropy}) one obtains
\begin{equation} \label{sente}
S_{BH} = 16\pi (e)^3
\end{equation}

Typically one is interested in expressing the results in terms of
charges, not field strengths, and this is achieved by using
(\ref{chgdef}). As shown in \cite{Castro:2007hc}, the results can be
put in compact form in the following way. We first define scaled 
moduli
\begin{equation}
\bar{M}^I \equiv \sqrt{v_1} M^I \;.
\end{equation}
Solution for them is implicitly given with
\begin{equation} \label{mbareq}
8\,c_{IJK} \bar{M}^J \bar{M}^K = q_I + \frac{c_{I}}{8}
\end{equation}
and the entropy (\ref{sente}) becomes
\begin{equation} \label{sentm}
S_{BH} = \frac{8\pi}{3} c_{IJK}\bar{M}^I \bar{M}^J \bar{M}^K 
\end{equation}
A virtue of this presentation is that if one is interested only in
entropy, then it is enough to consider just (\ref{mbareq}) and
(\ref{sentm}). It was shown in \cite{Castro:2007ci} that (\ref{sentm})
agrees with the OSV conjecture \cite{topstr,Guica:2005ig}, after proper
treatment of uplift from $D=4$ to $D=5$ is made.  

We shall be especially interested in the case when prepotential is
of the form
\begin{equation}
\mathcal{N} = \frac{1}{2}M^1c_{ij}M^iM^j \;, \qquad i,j>1
\end{equation}
where $c_{ij}$ is a regular matrix with an inverse $c^{ij}$.
In this case, which corresponds to $K3\times T^2$ 11-dimensional 
compactifications, it is easy to show that the entropy of BPS black 
holes is given with
\begin{equation} \label{seK3}
S_{BH} = 2\pi\sqrt{\frac{1}{2}|\hat{q}_1|c^{ij}\hat{q}_i\hat{q}_j} \;,
\qquad \hat{q_I}=q_I+\frac{c_I}{8}
\end{equation}

\section{$\mathcal{N} = M^1 M^2 M^3$ model -- heterotic 
 string on $T^4\times S^1$}
\label{sec:stu}

\subsection{BH solutions without corrections}
\label{ssec:sol0}

To analyse non-BPS solutions we take a simple model with $I=1,2,3$ and
prepotential
\begin{equation} \label{stupp}
\mathcal{N} = M^{1} M^{2} M^{3} \;,
\end{equation}
which is obtained when one compactifies 11-dimensional SUGRA on
six-dimensional torus $T^6$. It is known \cite{s11-het5,Ferrara:1996hh}
that with this choice one obtains tree level effective action of
heterotic string compactified on $T^4\times S^1$ which is
wounded around $S^1$. 

The simplest way to see this is to do the following steps. Start with
the Lagrangian in the on-shell form (\ref{l0noaux}), use (\ref{stupp})
with the condition $\mathcal{N}=1$, introduce two independent moduli
$S$ and $T$ such that
\begin{equation} \label{hetmod}
M^1 = S^{2/3} \;,\qquad M^2 = S^{-1/3}T^{-1} \;,\qquad M^3 = S^{-1/3}T
\end{equation}
Finally, make Poincar\'{e} duality transformation on the two-form gauge
field $F^1$: introduce additional 2-form $B$ with the 
corresponding strength $H=dB$ and add to the action a term
\begin{equation} \label{poidual}
\mathcal{A}_B = \frac{1}{4\pi^2} \int F^1 \land H 
= - \frac{1}{8\pi^2} \int dx^5 \sqrt{-g} F^1_{ab} (*H)^{ab}
\end{equation}
where $*$ is Hodge star. If one first solves for the $B$ field, the
above term just forces two-form $F^1$ to satisfy Bianchi identity, so 
the new action is classically equivalent to the starting one. But
if one solves for the $F^1$ and puts the solution back into the
action, after the dust settles one obtains that Lagrangian density
takes the form
\begin{equation} \label{lhetE}
4\pi^2\mathcal{L}_0 = R - \frac{1}{3}\frac{(\partial S)^2}{S^2}
 - \frac{(\partial T)^2}{T^2} 
 - \frac{S^{4/3}}{12}\left(H'_{abc}\right)^2 
 - \frac{1}{4}S^{2/3}T^2\left(F^2_{ab}\right)^2 
 - \frac{S^{2/3}}{4\,T^2}\left(F^3_{ab}\right)^2 
\end{equation}
where 3-form field $H'$ is defined with
\begin{equation} \label{lwh}
H'_{abc} = \partial_a B_{bc} - \frac{1}{2}
 \left( A^2_a F^3_{bc} + A^3_a F^2_{bc} \right) 
 + (\mbox{cyclic permutations of } a,b,c)
\end{equation}
To get the action in an even more familiar form one performs a Weyl
rescaling of the metric
\begin{equation}
g_{ab} \to S^{2/3} g_{ab}
\end{equation}
where in the new metric Lagrangian (\ref{lhetE}) takes the form
\begin{equation} \label{lhetS}
4\pi^2\mathcal{L}_0 = S \left[ R + \frac{(\partial S)^2}{S^2}
 - \frac{(\partial T)^2}{T^2} - \frac{1}{12}\left(H'_{abc}\right)^2
 - \frac{T^2}{4}\left(F^2_{ab}\right)^{\!2}
 - \frac{1}{4\,T^2}\left(F^3_{ab}\right)^{\!2} \right]
\end{equation} 

One can now check\footnote{For example by comparing with Eqs.\ (2.2),
(2.8) and (2.3) in \cite{Sen:2005kj}. Observe that, beside simple
change in indices $1\to2$ and $2\to3$, one needs to divide
gauge fields by a factor of two to get results in Sen's conventions.
There is also a difference in a convention for $\alpha'$, which makes
normalization of charges different.} that (\ref{lhetE})
and (\ref{lhetS}) are indeed lowest order (in $\alpha'$ and $g_s$) 
effective Lagrangians in Einstein and string frame, respectively,  of
the heterotic string compactified on $T^4\times S^1$ with the only 
"charges" coming from winding and momentum on $S^1$. Field $T$ plays
the role of a radius of $S^1$, and field $S$ is a function of a 
dilaton field such that $S\sim1/g_s^2$. This interpretation 
immediately forces all $M^I$ to be positive.

We are interested in finding 3-charge near-horizon solutions for BH's
when the prepotential is (\ref{stupp}). Applying entropy function
formalism on (\ref{f0susy}) one easily gets:
\begin{eqnarray}
v_1 &=& \frac{1}{4} \left|q_1q_2q_3\right|^{1/3} \label{v10sol} \\
e^I &=& \frac{4 v_1^{3/2}}{q_I} 
 = \frac{1}{2q_I} \left|q_1q_2q_3\right|^{1/2} \label{eI0sol} \\
M^I &=& \frac{|e^I|}{\sqrt{v_1}} 
 = \left|\frac{q_1q_2q_3}{q_I^2}\right|^{1/3} \label{merat} \\
v_2 &=& 4 v_1 \label{ansv21} \\
D &=& - \frac{1}{v_1} \left|\textrm{sign}(q_1)+\textrm{sign}(q_2)
 +\textrm{sign}(q_3)\right| \label{ansD} \\
V &=& \frac{\sqrt{v_1}}{4} \left(\textrm{sign}(q_1)
 +\textrm{sign}(q_2)+\textrm{sign}(q_3)\right) \label{ansV}
\end{eqnarray}
and the entropy is given with
\begin{equation} \label{stu0ent}
S = 2\pi \left|q_1q_2q_3\right|^{1/2}
\end{equation}
In fact in this case full solutions (not only near-horizon
but in the whole space) were explicitly constructed in 
\cite{Peet:1995pe}.

If any of charges $q_I$ vanishes, one gets singular solutions with
vanishing horizon area. Such solutions correspond to small black
holes. One expects that higher order (string) corrections ``blow up''
the horizon and make solutions regular.

\subsection{Inclusion of SUSY corrections}
\label{ssec:stusorr}

We would now like to find near horizon solutions for extremal black
holes when the action is extended with the supersymmetric higher
derivative correction (\ref{l1susy}). We already saw in 
Sec.\ \ref{sec:bps} how this can be done for the special case of 
1/2 BPS solutions, i.e., in case of non-negative charges $q_I\geq0$.
The question is could the same be done for general sets of charges.

Again, even for the simple prepotential (\ref{stupp}) any attempt of
direct solving of EOM's is futile. In the BPS case we used vanishing 
of all supersymmetry variations which gave the conditions
(\ref{merat})-(\ref{stu0ent}), which are not affected by higher
derivative correction, and that enabled us to find a complete 
solution. Now, for non-BPS case, we cannot use the same argument, and
naive guess that (\ref{merat}-\ref{stu0ent}) is preserved after
inclusion of correction is inconsistent with EOM's.

Intriguingly, there is something which is shared by (BPS and non-BPS)
solutions (\ref{v10sol})-(\ref{ansV}) -- the following two relations:
\begin{eqnarray} \label{anscond1}
0 &=& Dv_1 + 3 - 9\frac{v_1}{v_2} + 4\frac{V^2}{v_1} \label{asusy} \\
0 &=& \frac{(Dv_1)^2}{12} + 4\left(\frac{V}{\sqrt{v_1}}\right)^{\!4} 
 + \frac{1}{4}\left(1-\frac{v_1}{v_2}\right)^{\!2}
 - \frac{2}{3}\left(\frac{V}{\sqrt{v_1}}\right)^{\!2}
 \left(5+3\frac{v_1}{v_2}\right)
\label{anscond2}
\end{eqnarray}
The above conditions are connected with supersymmetry. The first one,
when plugged in the $\mathcal{L}_0$ (\ref{l0susy}), makes the first 
bracket (multiplying $\mathcal{A}^2$) to vanish. The second condition,
when plugged in the $\mathcal{L}_1$ (\ref{l1susy}), makes the term 
multiplying $c_{I}M^I$ to vanish. We shall return to this point in
Sec.\ \ref{sec:genpp}.

What is important is that for (\ref{anscond1}) and (\ref{anscond2}) we 
needed just Eqs.\ (\ref{ansv21})-(\ref{ansV}) (and, in particular, 
{\em not} Eq.\ (\ref{merat})). Our idea is to take 
(\ref{ansv21})-(\ref{ansV}) as an {\em ansatz}, plug this into all EOM's
and find out is it working also in the non-BPS case. Using the CPT 
symmetry it is obvious that there are just two independent cases. We can
choose
\begin{equation} \label{ans3}
v_2 = 4 v_1 \;, \qquad D = -\frac{3}{v_1} \;, \qquad
V = \frac{3}{4}\sqrt{v_1} \;,
\end{equation}
which corresponds to BPS case (see (\ref{bpscond})), and
\begin{equation} \label{ans1}
v_2 = 4 v_1 \;, \qquad D = -\frac{1}{v_1} \;, \qquad 
V = \frac{1}{4}\sqrt{v_1}
\end{equation}
Though in the lowest order (\ref{ans3}) appears when all charges are
positive, and (\ref{ans1}) when just one charge is negative 
(see (\ref{ansv21})-(\ref{ansV})), we shall {\em not} suppose a priori
any condition on the charges. 

For the start, let us restrict coefficients $c_{I}$ such that
\begin{equation} \label{c2het}
c_1 \equiv 24\zeta > 0 \;, \qquad c_2 = c_3 = 0 \;.
\end{equation}
This choice appears when one considers heterotic string effective
action on the tree level in string coupling $g_s$, but taking into
account (part of) corrections in $\alpha'$.\footnote{To consider 
corrections in $g_s$ it would be necessary also to make corrections 
in the prepotential (i.e., to $c_{IJK}$).} In this case we have 
$\zeta=1$. For completeness, we present results for general 
coefficients $c_{I}$ in Appendix \ref{sec:app1}. 

Let us now start with the ansatz (\ref{ans1}). The EOM's can now be 
written in the following form:
\begin{eqnarray}
&& b^2 b^3 e^2 e^3 = 0 
 \nonumber \\ 
&& b^1 b^3 e^1 e^3 = 0 
 \nonumber \\
&& b^1 b^2 e^1 e^2 = 0
 \nonumber \\
&& 4 \left(b^2 b^3 - 1\right) e^2 e^3 = q_1 - \frac{\zeta}{3}
 \nonumber \\
&& 4 \left(b^1 b^3 - 1\right) e^1 e^3 = q_2
 \nonumber \\
&& 4 \left(b^1 b^2 - 1\right) e^1 e^2 = q_3
 \nonumber \\
&& 42 v_1^{3/2} + \left(\zeta \left(6 b^1 - 1\right) +
 6 \left(4 b^1 b^2 b^3 - 3(b^1+1)(b^2+1)(b^3+1)+4\right)
 e^2 e^3\right) e^1 = 0
 \nonumber \\
&& 18 v_1^{3/2} + \left(
 6 \left(4 b^1 b^2 b^3 + (b^1+1)(b^2+1)(b^3+1)+4\right)
 e^2 e^3 - \zeta \left(2 b^1 + 5\right)\right) e^1 = 0
 \nonumber \\
&& 6 v_1^{3/2} + \left(\zeta(2 b^1 + 1) -
 6 \left(b^1 + 1\right) \left(b^2 + 1\right) \left(b^3 + 1\right)
 e^2 e^3\right) e^1 = 0
 \nonumber \\
&& 6 v_1^{3/2} + \left(\zeta \left(10 b^1 + 9\right)
 + 6 \left(3b^1b^2b^3 - b^1b^2- b^2b^3- b^1b^3 - 5(b^1+b^2+b^3)
 - 9\right) e^2 e^3\right) e^1 = 0
 \nonumber
\end{eqnarray}
where $b^I$ are defined with
\begin{equation}
\bar{M}^I \equiv (1+b^I)e^I
\end{equation}
Now there are more equations than unknowns, so the system is naively 
overdetermined. However, not all equations are independent and the
system is solvable. First notice that first three equations imply that
two of $b^I$'s should vanish, which enormously simplifies solving.

Let us summarize our results. We have found that there are six branches
of solutions satisfying\footnote{We note that, as was shown in $D=4$ 
\cite{Sen:2004}, that corrections can change relations between fields in
the action and moduli of the compactification manifold, so one should be 
careful when demanding physicality conditions.} $M^I>0$, depending on 
the value of the charges $q_I$.\\

\noindent
\underline{\emph{$q_1>\zeta/3$, $q_2>0$, $q_3<0$}}\\

\noindent
Solutions are given with:
\begin{eqnarray}
&& v_1 = \frac{1}{4}
 \left|\frac{q_2q_3(q_1+\zeta/3)^2}{q_1-\zeta/3}\right|^{1/3} \\
&& \frac{e^1}{\sqrt{v_1^3}}\left(q_1-\frac{\zeta}{3}\right)
 = \frac{e^2q_2}{\sqrt{v_1^3}} = \frac{e^3q_3}{\sqrt{v_1^3}}
 = 4\frac{q_1-\zeta/3}{q_1+\zeta/3}
\\
&& \frac{M^3\sqrt{v_1}}{e^3} = - \frac{q_1+\zeta}{q_1-\zeta/3}
 \;, \qquad \frac{M^1\sqrt{v_1}}{e^1}
 = \frac{M^2\sqrt{v_1}}{e^2} = 1
\end{eqnarray}
together with (\ref{ans1}). The entropy is given with
\begin{equation} \label{sent12}
S_{BH} = 2\pi\left|q_2q_3\left(q_1-\frac{\zeta}{3}\right)\right|^{1/2}
\end{equation}

For heterotic string one has $\zeta=1$ and $q_I$ are integer, so the
condition can be written also as $q_1>0$.\\

\noindent
\underline{\emph{$q_1>\zeta/3$, $q_2<0$, $q_3>0$}}\\

\noindent
As the theory is symmetric on the exchange $(2)\leftrightarrow(3)$,
the only difference from the previous case is that now we have
\begin{equation}
\frac{M^2\sqrt{v_1}}{e^2} = - \frac{q_1+\zeta}{q_1-\zeta/3}
 \;, \qquad \frac{M^1\sqrt{v_1}}{e^1}
 = \frac{M^3\sqrt{v_1}}{e^3} = 1
\end{equation}
and everything else is the same.\\

\noindent
\underline{\emph{$q_1<-\zeta$, $q_2>0$, $q_3>0$}}\\

\noindent
Here the only difference from solutions in previous two cases is:
\begin{equation}
\frac{M^1\sqrt{v_1}}{e^1} = - \frac{q_1-\zeta/3}{q_1+\zeta}
 \;, \qquad \frac{M^2\sqrt{v_1}}{e^2}
 = \frac{M^3\sqrt{v_1}}{e^3} = 1
\end{equation} 
For heterotic string $\zeta=1$ the bound for $q_1$ is $q_1<-1$.

Beside these three "normal" branches, there are additional three "strange"
branches which appear for $|q_1|<\zeta/3$:\\

\noindent
\underline{\emph{$|q_1|<\zeta/3$, $q_2<0$, $q_3<0$}}\\

\noindent
For every of the three branches discussed above, there is an additional,
mathematically connected, branch, for which the difference is that now
in all branches we have $|q_1|<\zeta/3$, $q_2<0$, $q_3<0$. All
formulas are the same, except that the entropy is negative
\begin{equation}
S_{BH} = -2\pi\left|q_2q_3\left(q_1-\frac{\zeta}{3}\right)\right|^{1/2}
\end{equation}
Additional reason why we call these solutions "strange" is the fact
that electric fields and charges have opposite sign. It is
questionable that there are asymptotically flat BH solutions with such
near-horizon behaviour, and for the rest of the paper we shall ignore
them.

\vspace*{2mm}
Now we take the ``BPS" ansatz (\ref{ans3}). There is only one branch of
solutions, valid for $q_{2,3}>0$, $q_1>-\zeta$:\\

\noindent
\underline{\emph{$q_1>-\zeta$, $q_2>0$, $q_3>0$}}\\

\noindent
Solution now takes the form
\begin{eqnarray}
&& v_1 = \frac{1}{4}
 \left|\frac{q_2q_3(q_1+\zeta)^2}{q_1+3\zeta}\right|^{1/3} \label{hetbpsv} \\
&& \frac{e^1}{\sqrt{v_1^3}}\left(q_1+3\zeta\right)
 = \frac{e^2q_2}{\sqrt{v_1^3}} = \frac{e^3q_3}{\sqrt{v_1^3}} 
 = 4\frac{q_1+3\zeta}{q_1+\zeta}
\label{hetbpse} \\
&& \frac{M^1\sqrt{v_1}}{e^1} 
 = \frac{M^2\sqrt{v_1}}{e^2}
 = \frac{M^3\sqrt{v_1}}{e^3} = 1
\label{hetbpsm}
\end{eqnarray}
together with (\ref{ans3}). The entropy is given with
\begin{equation} \label{sent34}
S_{BH} = 2\pi\left|q_2q_3\left(q_1+3\zeta\right)\right|^{1/2}
\end{equation}
One can check that this is equal to the BPS solution from 
Sec.\ \ref{sec:bps} with the prepotential and $c_I$ given by 
(\ref{stupp}) and (\ref{c2het}).

\vspace*{2mm}
Solutions for the cases when two or all three charges are negative are
simply obtained by applying the CPT transformations $e^I\to-e^I$, 
$q^I\to-q^I$, $V\to-V$ on the solutions above.

\subsection{Some remarks on the solutions}
\label{ssec:srem}

Let us summarize the results of Sec.\ 5.2. For the prepotential 
(\ref{stupp}) and (\ref{c2het}) we have found nonsingular extremal 
near-horizon solutions with $AdS_2\times S^3$ geometry for all values of 
charges $(q_1,q_2,q_3)$ except for some special cases. For black hole 
entropy we have obtained that supersymmetric higher order ($R^2$) 
correction just introduces a shift $q_1\to\hat{q}_1=q_1+a$,
\[
S_{BH} = 2\pi\sqrt{\left| \hat{q}_1 q_2 q_3 \right|}
\]
where $a=\pm3,\pm1/3$.

For the action connected with compactified heterotic string, i.e., when 
$\zeta=1$ and charges are integer valued, exceptions are:
\begin{description}
	\item [(i)] $q_2q_3=0$
	\item [(ii)] $q_1=0\;,\;\; q_2q_3<0$
	\item [(iii)] $q_1=-1\;,\;\; q_2,q_3>0$ (and also with reversed signs)
\end{description}
It is easy to show that in order to have small effective string coupling near the 
horizon we need $q_2q_3\gg1$ which precludes case (i) (string loop 
corrections make $c_{2,3}\neq0$ which regulate case (i), see 
Append. \ref{sec:app1}). For 
the cases (ii) and (iii) one possibility is that regular solutions exist, 
but they are not given by our Ans\"{a}tze. But, our efforts to find 
numerical solutions also failed, so it is also possible that such solutions
do not exist. This would not be that strange for cases (i) and (ii), as 
they correspond to black hole solutions which were already singular (small)
with vanishing entropy before inclusion of supersymmetric $R^2$ corrections.
But for the case (iii) it would be somewhat bizarre, because it would mean 
that higher order corrections turn nonsingular solution into singular.

Let us make a comment on a consequence of the violation of the real special 
geometry condition by supersymmetric higher-derivative corrections. We have seen
that the example analysed in this section can be viewed as the tree-level 
effective action of heterotic string compactified on $T^4\times S^1$ supplied with
part of $\alpha'$ corrections. In Sec.\ \ref{ssec:sol0} we saw that in the lowest 
order a radius $T$ of $S^1$ was identified with $T^2=M^3/M^2$. From 
(\ref{hetbpsv})-(\ref{hetbpsm}) follows that in the BPS solution we have
\begin{equation}
T^2 = \frac{q_2}{q_3}
\end{equation}
which is expected from T-duality $q_2\leftrightarrow q_3$, $T\to T^{-1}$.

But, in the lowest order we also have $T^2=M^1(M^3)^2$, which gives
 \begin{equation}
T^2 = \frac{q_2}{q_3} \frac{q_1+3}{q_1+1}
\end{equation}
which does not satisfy T-duality. It means that relation $T^2=M^1(M^3)^2$
receives higher-derivative corrections.\footnote{Similar observation in $D=4$
dimensions was given in \cite{Sen:2004}.} That at least one of relations for
$T$ is violated by corrections was of course expected from 
$\mathcal{N}\neq1$.\footnote{Notice that for some non-BPS solutions both 
relations are violated.}

\section{Generalisation to other prepotentials}
\label{sec:genpp}

A natural question would be to ask in what extend one can generalize
construction from the previous section. In mathematical terms, the
question is of validity of ansatz (\ref{ans1})
\begin{equation} \label{gans1}
v_2 = 4 v_1 \;, \qquad D = -\frac{1}{v_1} \;, \qquad
V = \frac{1}{4}\sqrt{v_1}
\end{equation}
which we call Ansatz 1, and (\ref{ans3})
\begin{equation} \label{gans3}
v_2 = 4 v_1 \;, \qquad D = -\frac{3}{v_1} \;, \qquad
V = \frac{3}{4}\sqrt{v_1} \;,
\end{equation}
which we call Ansatz 3 (Ansatz 2 and 4 are obtained by applying CPT
transformation, i.e, $V\to-V$).

We have seen in Sec.\ \ref{sec:bps} that for BPS states supersymmetry
directly dictates validity of Ansatz 3 (and by symmetry also 4). 
The remaining question is how general is Ansatz 1.

Putting (\ref{gans1}) in EOM's one gets
\begin{eqnarray} \label{sgeneom}
&& c_{IJK}e^{J}e^{K}+2 \bar{\mathcal{N}}_{I}=2
 \bar{\mathcal{N}}_{IJ}e^{J}
\nonumber \\ 
&& 6 \left({c_I \bar{M}^I}+168 v_1^{3/2}+24
\bar{\mathcal{N}}+48 \bar{\mathcal{N}}_{IJ}e^{I}e^{J}\right)=7 {c_I
e^I}+576 \bar{\mathcal{N}}_{I}e^{I}
\nonumber \\
&& 144 \left(3 v_1^{3/2}+5 \bar{\mathcal{N}}+2
\bar{\mathcal{N}}_{IJ}e^{I}e^{J}\right)=3 {c_I e^I}+2 {c_I
\bar{M}^I}+576 \bar{\mathcal{N}}_{I}e^{I}
\nonumber \\ 
&& {c_I e^I}+144 \bar{\mathcal{N}} = 2({c_I\bar{M}^I}+72 v_1^{3/2})
\nonumber \\
&& {c_I e^I}+576 \bar{\mathcal{N}}_{I}e^{I}=10 {c_I \bar{M}^I}+144
 v_1^{3/2}+432 \bar{\mathcal{N}}
\nonumber \\
&& q_I - \frac{c_I}{72} = 4\bar{\mathcal{N}}_{I} 
 - 4\bar{\mathcal{N}}_{IJ}e^{J}
\end{eqnarray}
and for the black hole entropy
\begin{equation}
S_{BH} = 4\pi
\left( 2\bar{\mathcal{N}}-\bar{\mathcal{N}}_{IJ}e^{I}e^{J} \right)
 = \frac{4\pi}{3}\hat{q}_Ie^I
\end{equation}
It can be shown that two equations in (\ref{sgeneom}) are not 
independent. In fact, by further manipulation the system can be put in 
the simpler form
\begin{eqnarray}
&& 0 = c_{IJK}\left(\bar{M}^J-e^J\right)\left(\bar{M}^K-e^K\right)
\label{a1eom1} \\
&& \frac{c_I\bar{M}^I}{12} = c_{IJK} \left(\bar{M}^I+e^I\right)
 \bar{M}^J e^K
\label{a1eom3} \\
&& v_1^{3/2} = \frac{c_Ie^I}{144} - (e)^3
\label{a1eom4} \\
&& q_I - \frac{c_I}{72} = -2\,c_{IJK}e^{J}e^{K}
\label{a1eom5} 
\end{eqnarray}
Still the above system is generically overdetermined as there is one 
equation more than the number of unknowns. More precisely, Eqs.\ 
(\ref{a1eom1}) and (\ref{a1eom3}) should be compatible, and this is
not happening for generic choice of parameters. One can check this,
e.g., by numerically solving simultaneously (\ref{a1eom1}) and 
(\ref{a1eom3}) for random choices of $c_{IJK}$, $c_I$ and $e^I$. This 
means that for generic prepotentials the Ansatz 1 (\ref{gans1}) is not
working.

However, there are cases in which the system is regular and there are
physical solutions. This happens, e.g., for prepotentials of the type
\begin{equation}
\mathcal{N} = \frac{1}{2}M^1c_{ij}M^iM^j \;, \qquad i,j>1
\end{equation}
where $c_{ij}$ is a regular matrix.
In this case (\ref{a1eom1}) gives conditions
\begin{equation}
0 = \left(\bar{M}^1-e^1\right)\left(\bar{M}^i-e^i\right) \;, \qquad
0 = \left(\bar{M}^i-e^i\right)c_{ij}\left(\bar{M}^j-e^j\right)
\end{equation}
which has one obvious solution when $\bar{M}^i=e^i$ for all $i$. Now
$\bar{M}^1$ is left undetermined, and one uses ``the extra equation''
(\ref{a1eom3}) to get it. Black hole entropy becomes
\begin{equation}
S_{BH} = 2\pi\sqrt{\frac{1}{2}|\hat{q}_1|c^{ij}\hat{q}_i\hat{q}_j} \;,
\qquad \hat{q_I}=q_I-\frac{c_I}{72}
\end{equation}
where $c^{ij}$ is matrix inverse of $c_{ij}$. Again, the influence of
higher order supersymmetric correction is just to shift
electric charges $q_I\to\hat{q}_I$, but with the different value for
the shift constant than for BPS black holes.

We have noted in Sec.\ \ref{ssec:stusorr} that Ansatz 1 (\ref{ans1}), 
which gives nonsupersymmetric solutions, has some interesting relations 
with supersymmetry. Another way to see this is to analyse
supersymmetry variations (\ref{svar}). Let us take that spinor
parameters $\eta$ and $\varepsilon$ are now connected with
\begin{equation}
\eta^i = (\gamma\cdot v)\varepsilon^i
\end{equation}
The variations (\ref{svar}) now become
\begin{eqnarray} \label{svarn1}
\delta\psi_\mu^i &=& \left(\mathcal{D}_\mu
 + \frac{1}{2}v^{ab}\gamma_{\mu ab}
 - \gamma_\mu(\gamma\cdot v)\right)\varepsilon^i \\
\delta\xi^i &=& \left(D + 4(\gamma\cdot v)^2\right)\varepsilon^i 
 \label{svarn2} \\
\delta\Omega^{Ii} &=& - \left(\frac{1}{4}\gamma\cdot F^{I}
 + M^{I}\gamma\cdot v \right)\varepsilon^i
 \label{svarn3} \\
\delta\zeta^{\alpha} &=&
 2(\gamma\cdot v)\varepsilon^j\mathcal{A}_j^\alpha
 \label{svarn4}
\end{eqnarray}
One can take a gauge in which
$\mathcal{A}_j^\alpha=\delta_j^\alpha$, which means that last
(hypermultiplet) variation (\ref{svarn4}) is now nonvanishing. But, it
is easy to see that for Ansatz 1 (and when $\epsilon^i$ is Killing 
spinor) variations (\ref{svarn1}) and (\ref{svarn2}) are vanishing. 
Also, we have seen that solutions we have been explicitly constructed
have the property that for all values of the index $I$ 
{\em except one} (which we denote $J$) we had
\begin{equation}
\bar{M}^I=e^I \; \qquad I\neq J
\end{equation}
From this follows that all variations (\ref{svarn3}) except the one for
$I=J$ are also vanishing. One possible explanation for such partial 
vanishing of variations could be that our non-BPS states of $N=2$ SUGRA are 
connected with BPS states of some theory with higher (e.g., $N=4$) 
supersymmetry.

\section{Gauss-Bonnet correction}
\label{sec:GB}

It is known that in some cases of black holes in $D=4$ Gauss-Bonnet term 
somehow effectively takes into account all $\alpha'$ string
corrections. Let us now investigate what is happening in $D=5$. This
means that we now add as $R^2$ correction to the 0$^{th}$ order 
Lagrangian (\ref{l0susy}) instead of (\ref{l1susy}) just the term 
proportional to the Gauss-Bonnet density:
\begin{equation} \label{l1gb}
\mathcal{L}_{GB} = \frac{1}{4\pi^2}\frac{1}{8}\frac{c_{I}M^I}{24}
 \left(R_{abcd}R^{abcd}-4R_{ab}R^{ab}+R^2\right)
\end{equation}
To apply entropy function formalism we start with
\begin{equation} \label{fGB}
f = f_0 + f_{GB}
\end{equation}
where $f_0$ is again given in (\ref{f0susy}) and $f_{GB}$ is
\begin{equation} \label{f1GB}
f_{GB} = -\frac{3}{2}\sqrt{v_2}\frac{c_{I}M^I}{24}
\end{equation}

Strictly speaking, we have taken just (part of) first order correction
in $\alpha'$, so normally we would expect the above action to give us at
best just the first order correction in entropy. This we obtain by putting
$0^{th}$-order solution in the expression
\begin{equation} \label{delS}
\Delta S_{BH} = -2\pi \Delta f
\end{equation}
where $\Delta f$ is 1$^{st}$-order correction in $f$. It is easy to
show that for the BPS $0^{th}$-order solution (\ref{bpscond}) one 
obtains the same result for supersymmetric (\ref{f1susy}) and 
Gauss-Bonnet (\ref{f1GB}) corrections, which can be written in a form:
\begin{equation} \label{ealpha}
\Delta S_{BH} = 6\pi\frac{c_{I}e^I}{24}
\end{equation}
It was noted in \cite{Castro:2007hc} that for compactifications on 
elliptically fibred Calabi-Yau (\ref{ealpha}) agrees with 
the correction of microscopic entropy proposed earlier by Vafa
\cite{Vafa:1997gr}. We note that for non-BPS black holes already
first $\alpha'$ correction to entropy is different for SUSY and
Gauss-Bonnet case.

From experience in $D=4$ one could be tempted to suppose that SUSY and
Gauss-Bonnet solutions are exactly (not just perturbatively) equal.
However, this is not true anymore in $D=5$. The simplest way to see
this is to analyse opposite extreme where one of the charges is zero
(small black holes). To explicitly show the difference let us analyse
models of the type (obtained from $K3\times T^2$ compactifications of 
$D=11$ SUGRA)
\begin{equation}
\mathcal{N}=\frac{1}{2}M^1c_{ij}M^iM^j \;, \qquad c_i=0 \;, \qquad
 i,j>1
\end{equation}
in the case where $q_1=0$. For the Gauss-Bonnet correction,
application of entropy function formalism of Sec.\ \ref{sec:ent-func}
on (\ref{fGB}) gives for the entropy (see Appendix \ref{sec:app2})
\begin{equation} \label{eGBK3}
S_{GB} = 4\pi\sqrt{\frac{1}{2}\frac{c_1}{24}q_ic^{ij}q_j}
\end{equation}
where $c^{ij}$ is the matrix inverse of $c_{ij}$. On the other hand, 
from (\ref{seK3}) follows that for the supersymmetric correction in
the BPS case one gets
\begin{equation} \label{esK3}
S_{SUSY} = 2\pi\sqrt{\frac{3}{2}\frac{c_1}{24}q_ic^{ij}q_j}
\end{equation}
which is differing from (\ref{eGBK3}) by a factor of $2/\sqrt{3}$.

In \cite{Huang:2007sb} some of the models of this type were analysed 
from microscopic point of view and the obtained entropy of small black
holes agrees with the Gauss-Bonnet result (\ref{eGBK3}).

Now, the fact that simple Gauss-Bonnet correction is giving the
correct results for BPS black hole entropy in both extremes, $q_1=0$
and $q_1>>1$, is enough to wonder could it be that it gives the 
correct microscopic entropy for all $q_1\geq0$ (as it gives for 4 and
8-charge black holes in $D=4$). Analytical results, with details of
calculation, for the generic matrix $c_{ij}$ and charge $q_3$ are
presented in Appendix \ref{sec:app2}.

Here we shall present results for the specific case, already mentioned
in Sec.\ \ref{sec:stu}, of the heterotic string compactified on
$T^4\times S^1$. Tree-level (in $g_s$) effective action is defined with
\begin{equation}
\mathcal{N} = M^1 M^2 M^3 \;,\qquad c_1 = 24 \;,\qquad c_2 = c_3 =0 \;.
\end{equation}
Matrix $c_{ij}$ is obviously here given with
\begin{equation}
c_{12} = c_{21} = 1 \;, \qquad c_{11} = c_{22} = 0
\end{equation}
As the simple Gauss-Bonnet correction (\ref{l1gb}) does not contain 
auxiliary fields, we can integrate them out in the same way as it was
done in the lowest-order case in Sec.\ \ref{ssec:sol0}. For independent
moduli we again use
\begin{equation}
S \equiv (M^1)^{3/2} \;\qquad T \equiv \tilde{M}^2 = S^{1/3} M^2
\end{equation}
It appears that it is easier to work in string frame, where the 
0$^{th}$ order action is given in (\ref{lhetS}), and the correction 
(\ref{l1gb}) is now
\begin{equation} \label{lgbS}
\mathcal{L}_{GB} = \frac{1}{4\pi^2}\frac{S}{8}
 \left(R_{abcd}R^{abcd}-4R_{ab}R^{ab}+R^2\right) + 
 \mbox{ (terms containing } \partial_a S \mbox{)}
\end{equation}
We are going to be interested in near-horizon region where all 
covariant derivatives, including $\partial_a S$, vanish, so we can
again just keep Gauss-Bonnet density term.

Application of (\ref{efgen}) here gives that solution
near the horizon has the form
\begin{eqnarray} \label{efhgb}
&& ds^2 = v_1 \left( -x^2 dt^2 + \frac{dx^2}{x^2} \right)
 + v_2\,d\Omega_{3}^2 \nonumber \\
&& S(x) = S \;, \qquad T(x) = T \nonumber \\
&& F^{(i)}_{tr}(x) = -e_i \;,\qquad i=2,3  \nonumber \\
&& H_{mnr} = 2 q_1 \sqrt{h_S}\, \varepsilon_{mnr}
\end{eqnarray}
where $\varepsilon_{mnr}$ is totally antisymmetric tensor with
$\varepsilon_{234}=1$. Observe that $q_1$ is now a magnetic charge.
Using this in (\ref{lhetS}) and (\ref{lgbS}) gives
\begin{equation}
f = \frac{1}{2} v_1 v_2^{3/2} S \left( -\frac{2}{v_1} +
 \frac{6}{v_2} + \frac{T^2 e_2^2}{2 v_1^2} + \frac{e_3^2}{2 T^2 v_1^2}
 - \frac{2 q_1^2}{v_2^3} - \frac{3}{v_1 v_2} \right)
\end{equation}
Following the entropy function formalism we need to solve the system of 
equations
\begin{equation} \label{gbheteom}
0 = \frac{\partial f}{\partial v_1} \;, \qquad
0 = \frac{\partial f}{\partial v_2} \;, \qquad
0 = \frac{\partial f}{\partial S} \;, \qquad
0 = \frac{\partial f}{\partial T} \;, \qquad
q_2 = \frac{\partial f}{\partial e_2} \;, \qquad
q_3 = \frac{\partial f}{\partial e_3}
\end{equation}
After some straightforward algebra we obtain
\begin{equation}
T^2 = \left|\frac{q_2}{q_3}\right|
\end{equation}
which is the same as without the correction and respecting T-duality.
Also
\begin{equation}
v_1 = \frac{v_2}{4}+\frac{1}{8} \;, \qquad
S = \frac{1}{v_2}\sqrt{\frac{2v_2+1}{2v_2+3}}\sqrt{|q_2q_3|} \;.
\end{equation}
Here $v_2$ is the real root of a cubic equation
\begin{equation}
0 = x^3 - \frac{3}{2} x^2 - q_1^2 x - \frac{q_1^2}{2}
\end{equation}
which, explicitly written, is
\begin{eqnarray}
v_2 &=& \frac{1}{2}+\frac{(1+i\sqrt{3}) (4q_1^2+3)}{4\,3^{1/3}
 \left(-9-36q_1^2+2\sqrt{3}\sqrt{27 q_1^2+72 q_1^4-16 q_1^6}
 \right)^{1/3}}
\nonumber \\
&& +\frac{(1-i\sqrt{3})\left(-9-36q_1^2+2\sqrt{3}
 \sqrt{27q_1^2+72 q_1^4-16q_1^6}\right)^{1/3}}{4\,3^{2/3}}
\end{eqnarray}
For the macroscopic black hole entropy we obtain
\begin{equation} \label{GBhet}
S_{BH} = 4\pi\sqrt{|q_2q_3|}\sqrt{v_1+\frac{3}{2}\frac{v_1}{v_2}}
\end{equation}
It would be interesting to compare this result with the statistical
entropy of BPS states (correspondingly charged) in heterotic string 
theory. Unfortunately, this result is still not known.

For small 2-charge black holes $q_1=0$, and the solution further 
simplifies to
\begin{equation}
v_1 = \frac{v_2}{3} = \frac{1}{2}
\end{equation}
which gives for the entropy of small black holes
\begin{equation}
S_{BH} = 4\pi\sqrt{|q_2q_3|}
\end{equation}
This solution was already obtained in \cite{Sen:2005kj} by starting at
the beginning with $q_1=0$.\footnote{Notice that we are using 
$\alpha'=1$ convention, and in \cite{Sen:2005kj} it is $\alpha'=16$.
One can use the results from \cite{Prester:2005qs} to make connection
between conventions.}

\section{Conclusion and outlook}
\label{sec:concl}

We have shown that for some prepotentials, including important family
obtained with $K3\times T^2$ compactifications of 11-dimensional SUGRA,
one can find non-BPS spherically symmetric extremal black hole near 
horizon solutions. In particular, for the simple example of so called 
$STU$ theory we have explicitly constructed solutions for all values of
charges with the exception of some small black holes where one of the
charges is equal to 0 or $\pm1$.

One of the ideas was to compare results with the ones obtained by taking
$R^2$ correction to be just given with Gauss-Bonnet density, and 
especially to analyse cases when the actions are connected with 
string compactifications, like e.g., heterotic string on $K3\times S^1$,
where for some instances one can find statistical entropies. Though for 
Gauss-Bonnet correction (which manifestly breaks SUSY) it was not 
possible to calculate entropy in a closed form for generic prepotentials,
on some examples we have explicitly shown that in $D=5$, contrary to 
$D=4$ examples, black hole entropy is different from the one obtained 
using supersymmetric correction (BPS or non-BPS case). Interestingly, 
first order corrections to entropy of BPS black holes are the same for 
all prepotentials, and are in agreement with the result for statistical 
entropy for elliptically fibred Calabi-Yau compactification 
\cite{Vafa:1997gr}.

For the $K3\times T^2$ compactifications of $D=11$ SUGRA (which includes
$K3\times S^1$ compactification of heterotic string) we have found 
explicit formula for the black hole entropy in the case of Gauss-Bonnet
correction. Unfortunately, expression for statistical entropy for generic
values of charges is still not known, but there are examples for which
statistical entropy of BPS states corresponding to {\em small} black 
holes is known \cite{Huang:2007sb}. We have obtained that Gauss-Bonnet 
correction leads to the macroscopic entropy equal to statistical, contrary 
to supersymmetric correction which leads to different result. This
result favors Gauss-Bonnet correction. On the other hand, for 
{\em large} black holes, it is the supersymmetric result (\ref{seK3}) 
which agrees with OSV conjecture properly uplifted to $D=5$
\cite{Castro:2007ci}. We propose to resolve this issue perturbatively
by calculating $\alpha'^2$ correction for 3-charge black holes in 
heterotic string theory compactified on $K3\times S^1$ using methods 
of \cite{Sahoo:2006pm}. Calculation is underway and results will be 
presented elsewhere \cite{Cvitan:2007hu}.

It is known that theories in which higher curvature correction are given by
(extended) Gauss-Bonnet densities have special properties, some of which are
unique. Beside familiar ones (equations of motion are ``normal'' second 
order, in flat space and some other backgrounds they are free of ghosts and 
other spurious states, have well defined boundary terms and variational 
problem, first and second order formalisms are classically equivalent, 
extended Gauss-Bonnet densities have topological origin and are related to 
anomalies, etc), they also appear special in the approaches where black hole 
horizon is treated as a boundary and entropy is a consequence of the broken 
diffeomorphisms by the boundary condition \cite{CPP}. It would be interesting
to understand in which way this is connected with the observed fact
that these terms effectively encode a lot of near-horizon properties for a 
class of BPS black holes in string theory.

\acknowledgments

We would like to thank L. Bonora for stimulating discussions.
This work was supported by the Croatian Ministry of Science,
Education and Sport under the contract no. 119-0982930-1016. P.D.P.
was also supported by Alexander von Humboldt Foundation.

\appendix

\section{Solutions for $\mathcal{N}=M^1M^2M^3$ but general $c_I$}
\label{sec:app1}

In this appendix we consider actions with
\begin{equation}
\mathcal{N}=M^1M^2M^3
\end{equation}
but arbitrary coefficients $c_I$. Let us define
\begin{equation}
\zeta_I \equiv \frac{c_I}{24}
\end{equation}
and for simplicity restrict to $\zeta_I > 0$. We shall concentrate on
non-BPS solutions and Ansatz 1. For this case we can specialize the
general expression for relation between electric charges and field
strengths (\ref{a1eom5}) as follows
\begin{equation}
\hat{q}_{1} = -4 e^{2} e^{3} \ , \ \hat{q}_{2} = -4 e^{3} e^{1} \ , 
 \ \hat{q}_{3} = -4 e^{1} e^{2}\\
\end{equation}
where we introduced shifted charges
\begin{equation}
\hat{q}_{I} \equiv q_I - \frac{\zeta_I}{3}
\end{equation}
From here follow also simple relations
\begin{equation}
\frac{e^{1} \hat{q}_{1}}{4} = \frac{e^{2} \hat{q}_{2}}{4}
 = \frac{e^{3} \hat{q}_{3}}{4} = -(e)^{3}
\end{equation}
We introduce definition
\begin{equation} \label{amev}
A_{i} = \frac{M^{i}}{e^{i}} \sqrt{v_{1}} \ , \quad i=1,2,3 
\end{equation}
The corresponding system of equations then follows from equation
(\ref{sgeneom})
\begin{eqnarray}
&& (-1 + A_{2})(-1 + A_{3})e^{2}e^{3} = 0\\
&& (-1 + A_{1})(-1 + A_{3})e^{1}e^{3} = 0\\
&& (-1 + A_{1})(-1 + A_{2})e^{1}e^{2} = 0
\end{eqnarray}
\begin{eqnarray}
&& 6(7v_{1}^{3/2} + A_{1}e^{1}((4 - 4A_{2} + (-4 + A_{2})A_{3})e^{2}e^{3} + \zeta_{1}) + \nonumber\\ 
&& + A_{2}e^{2}(-4(-1 + A_{3})e^{1}e^{3} + \zeta_{2}) + A_{3}e^{3}(4e^{1}e^{2} + \zeta_{3})) = 7(e^{1}\zeta_{1} + e^{2}\zeta_{2} + e^{3}\zeta_{3})\\ 
&& 18v_{1}^{3/2} + 2e^{1}(3(4A_{3} + A_{1}(4 - 4A_{3} + A_{2}(-4 + 5A_{3})))e^{2}e^{3} - A_{1}\zeta_{1}) = \nonumber\\
&& = 3e^{1}\zeta_{1} + 3e^{2}\zeta_{2} + 2A_{2}e^{2}(12(-1 + A_{3})e^{1}e^{3} + \zeta_{2}) + (3 + 2A_{3})e^{3}\zeta_{3}\\
&& 6v_{1}^{3/2} + 2A_{1}e^{1}(-3A_{2}A_{3}e^{2}e^{3} + \zeta_{1}) + (-1 + A_{2})e^{2}\zeta_{2} + (-1 + A_{3})e^{3}\zeta_{3} = e^{1}\zeta_{1}\\
&& 6v_{1}^{3/2} + 2e^{1}(3(-4A_{2}A_{3} + A_{1}(-4A_{3} + A_{2}(-4 + 3A_{3})))e^{2}e^{3} + 5A_{1}\zeta_{1}) + \nonumber\\
&& + 10A_{2}e^{2}\zeta_{2} + (-1 + 10A_{3})e^{3}\zeta_{3} = e^{1}\zeta_{1} + e^{2}\zeta_{2}
\end{eqnarray}

We shall again find solutions with one negative and two positive
shifted charges, and ``strange'' solutions with all shifted charges negative.\\

\noindent
\underline{\emph{$\hat{q}_{1} < -4\zeta_{1}/3 \;,\quad \hat{q}_{2} > 0\;,
\quad \hat{q}_{3} > 0$}}\\

\noindent
We first describe solutions with one charge negative, e.~g., $q_1$.
Then
\begin{equation}\label{v1qz}
\sqrt{v_{1}} = \frac{1}{2} \,
 \frac{(-Q_{(3)}^{2})^{1/6}}{\hat{q}_{1}^{1/6}\hat{q}_{2}^{1/6}
 \hat{q}_{3}^{1/6}}
\end{equation}
\begin{equation}
Q_{(3)} \equiv \hat{q}_{1}\hat{q}_{2}\hat{q}_{3} + \frac{2}{3} \,
 \left(\zeta_{1}\hat{q}_{2}\hat{q}_{3}
 + \zeta_{2}\hat{q}_{1}\hat{q}_{3}
 + \zeta_{3}\hat{q}_{1}\hat{q}_{2}\right)
\end{equation}
\begin{equation}\label{aqz}
A_{1} = - \frac{(4\zeta_{3}\hat{q}_{2} + (4\zeta_{2} + 3q_{2})
 \hat{q}_{3})\hat{q}_{1}}{(4\zeta_{1}
 + 3\hat{q}_{1})\hat{q}_{2}\hat{q}_{3}} \;,\qquad  A_{2} = 1
 \;,\qquad  A_{3} = 1
\end{equation}
\begin{equation}
M^{1} = \frac{1}{48v_1^{2}} \, \frac{Q_{(3)}}{\hat{q}_{2}\hat{q}_{3}}
 \,  \frac{4\zeta_{3}\hat{q}_{2} + (4\zeta_{2} + 3\hat{q}_{2})
 \hat{q}_{3}}{\hat{q}_{1} + \frac{4}{3}\zeta_{1}}
\end{equation}
\begin{equation}\label{m23qz}
M^{2} = 4v_1\frac{\hat{q}_{1}\hat{q}_{3}}{Q_{(3)}} \;,\qquad 
 M^{3} = 4v_1\frac{\hat{q}_{1}\hat{q}_{2}}{Q_{(3)}}
\end{equation}
Here we are able to impose positivity restriction,
\begin{equation}
M^{i} > 0 \ , \ i=1,2,3
\end{equation}
\begin{equation}
M^{2} > 0 \Rightarrow \frac{\hat{q}_{3}}{Q_{(3)}} < 0 \Rightarrow Q_{(3)} < 0
\end{equation}
$M^{3} > 0$ is automatically satisfied.\\
Consider now $M^{1} > 0$. Note first that $Q_{3}/\hat{q}_{2}\hat{q}_{3} < 0$. Thus we obtain that
\begin{equation}
\frac{4\zeta_{3}\hat{q}_{2} + (4\zeta_{2} + 3\hat{q}_{2})\hat{q}_{3}}{\hat{q}_{1} + \frac{4}{3}\zeta_{1}} < 0
\end{equation}
But numerator is positive so we get
\begin{equation}
\hat{q}_{1} + \frac{4}{3}\zeta_{1} < 0
\end{equation}
Note that under mentioned restrictions the property $Q_{(3)} < 0$ is indeed satisfied. In fact
\begin{eqnarray}
Q_{(3)} &=& \hat{q}_{1}\hat{q}_{2}\hat{q}_{3}
 + \frac{2}{3}(\zeta_{1}\hat{q}_{2}\hat{q}_{3}
 + \zeta_{2}\hat{q}_{1}\hat{q}_{3} + \zeta_{3}\hat{q}_{1}\hat{q}_{2})
\nonumber\\
&<& - \frac{4}{3} \, \zeta_{1}\hat{q}_{2}\hat{q}_{3}
 + \frac{2}{3} \, (\zeta_{1}\hat{q}_{2}\hat{q}_{3}
 + \zeta_{2}\hat{q}_{1}\hat{q}_{3} + \zeta_{3}\hat{q}_{1}\hat{q}_{2}) 
\nonumber\\
&=& - \frac{2}{3} \, \zeta_{1}\hat{q}_{2}\hat{q}_{3}
 + \frac{2}{3} \, (\zeta_{2}\hat{q}_{1}\hat{q}_{3}
 + \zeta_{3}\hat{q}_{1}\hat{q}_{2}) < 0
\end{eqnarray}
Let us find entropy. We shall use (\ref{entropy}), (\ref{fsusy}) and
(\ref{amev}) to obtain:
\begin{equation}
F = 8\pi(e)^{3}\left\{ A_{1}A_{2}A_{3} - (A_{1} + A_{2} + A_{3}) \right\}
\end{equation}
But $A_{2} = A_{3} = 1$ so
\begin{equation}
S_{BH} = -16\pi(e)^{3}
\end{equation}
From the explicit form of the solution (\ref{v1qz})-(\ref{m23qz}) we have
\begin{equation}
S_{BH} = 2\pi \, 
 \textrm{sign}(\frac{\hat{q}_{1}\hat{q}_{2}\hat{q}_{3}}{Q_{(3)}})
 \sqrt{-\hat{q}_{1}\hat{q}_{2}\hat{q}_{3}}
\end{equation}
But $\textrm{sign}(\hat{q}_{1}\hat{q}_{2}\hat{q}_{3}/Q_{(3)}) = +1$ so
\begin{equation}
S_{BH} = 2\pi \sqrt{|\hat{q}_{1}\hat{q}_{2}\hat{q}_{3}|}
\end{equation}
We finally conclude that presented solution is valid for
\begin{equation}
\hat{q}_{3} > 0 \ , \ \hat{q}_{1} < -\frac{4}{3} \zeta_{1} \ , \ \hat{q}_{2} > 0
\end{equation}
completely analogous to the first case in Sec.\ \ref{ssec:stusorr}.\\

\noindent
\underline{\emph{$\hat{q}_{1}>0 \ , \ \hat{q}_{2}<-\frac{4}{3} \zeta_{2} \ , \ 
 \hat{q}_{3}>0$}}\\

\noindent
In this case all relations can be obtained from previous case with exchange 
$(1) \leftrightarrow (2)$.\\

\noindent
\underline{\emph{$\hat{q}_{1}>0 \ , \ \hat{q}_{2}>0 \ , \ 
 \hat{q}_{3}<-\frac{4}{3} \zeta_{3}$}}\\

\noindent
Analogously this case can be obtained from the first case with interchange 
$(1) \leftrightarrow (3)$.\\

In addition to described 3 ``normal'' branches there are also 3 ``strange'' 
branches which give negative entropy:
\begin{equation}
S_{BH} = -2\pi \sqrt{\left|\hat{q}_{1}\hat{q}_{2}\hat{q}_{3}\right|}
\end{equation}
Such a solution may occur only if all $\hat{q}_{I}$'s are negative.



\section{Gauss-Bonnet correction in $K3\times T^2$ compactifications}
\label{sec:app2}

In this appendix we give the proof of the Eqs.\ (\ref{eGBK3}) and 
(\ref{GBhet}). We consider the actions with
\begin{equation}
\mathcal{N}=\frac{1}{2}M^1c_{ij}M^iM^j \;, \qquad c_1 \equiv 24\zeta
 \;,\qquad c_i=0 \;, \qquad
 i,j>1
\end{equation}
and when the higher order correction is proportional to the
Gauss-Bonnet density, i.e., it is given with (\ref{l1gb}). For such 
corrections one can integrate out auxiliary fields in the same manner as
when there is no correction, and pass to the on-shell form of the action
which is now given with (\ref{l0noaux}) and (\ref{l1gb}), with the 
condition for real special geometry $\mathcal{N}=1$ (which is here 
{\em not} violated by higher order Gauss-Bonnet corrections) implicitly
understood.

Now, before going to a hard work, it is convenient to do following 
transformations (which is a generalisation of what we did in 
Sec.\ \ref{ssec:sol0}). First we introduce scaled moduli $\bar{M}^i$ 
and the dilaton $S$
\begin{equation}
S = (M^1)^{3/2} \;\qquad \tilde{M}^i = S^{1/3} M^i
\end{equation}
for which the real special geometry condition now reads
\begin{equation} \label{cmm1}
\frac{1}{2} c_{ij} \tilde{M}^i \tilde{M}^j = 1
\end{equation}
This condition fixes one of $\tilde{M}^i${\'{}}s. Then we make Poincar\'{e} 
duality transformation (\ref{poidual}) which replaces two-form gauge field 
strength $F^1$ with its dual 3-form strength $H$. 
Finally, we pass to the string frame metric by Weyl rescaling
\begin{equation}
g_{ab} \to S^{2/3} g_{ab} \;.
\end{equation}
Again, we are interested in $AdS_2\times S^3$ backgrounds which in the
present case requires
\begin{eqnarray}
&& ds^2 = v_1 \left( -x^2 dt^2 + \frac{dx^2}{x^2} \right)
 + v_2\,d\Omega_{3}^2 \nonumber \\
&& F^{i}_{tr}(x) = -e^i \;, \qquad
 H_{mnr} = 2 q_1 \sqrt{h_S}\,\varepsilon_{mnr} \nonumber \\
&& \tilde{M}^i(x) = \tilde{M}^i \;, \qquad S(x) = S
\end{eqnarray}
where $\varepsilon_{mnr}$ is totally antisymmetric tensor satisfying
$\varepsilon_{234}=1$. Observe that $q_1$ now plays the role of 
{\em magnetic} charge. We apply entropy function formalism of 
Sec.\ \ref{sec:ent-func}. Function $f$ is now
\begin{equation} \label{fK3gb}
f = \frac{1}{2}v_1v_2^{3/2}S \left( -\frac{2}{v_1} + \frac{6}{v_2}
 + \frac{1}{v_1^2}\tilde{G}_{ij}e^ie^j - \frac{2q_1^2}{v_2^3}
 - \frac{3\zeta}{v_1v_2} \right)
\end{equation}
where $\tilde{G}_{ij}$ is given with
\begin{equation} \label{gijK3}
\tilde{G}_{ij} = \frac{1}{2}\left( c_{ik}\tilde{M}^kc_{jl}\tilde{M}^l
 - c_{ij} \right)
\end{equation}
To obtain solutions we need to solve extremization equations
\begin{equation} \label{gbeom}
0 = \frac{\partial f}{\partial v_1} \;, \qquad
0 = \frac{\partial f}{\partial v_2} \;, \qquad
0 = \frac{\partial f}{\partial S} \;, \qquad
0 = \frac{\partial f}{\partial \tilde{M}^i} \;, \qquad
q_i = \frac{\partial f}{\partial e^i}
\end{equation}
From the third equation (for $S$) one obtains that $f$ is vanishing.
This allows us to solve immediately first two equations
(for $v_1$ and $v_2$) and obtain
\begin{equation} \label{gbv1}
v_1 = \frac{v_2}{4}+\frac{\zeta}{8}
\end{equation}
where $v_2$ is the real positive root of a cubic equation
\begin{equation}
0 = x^3 - \frac{3}{2}\zeta x^2 - q_1^2 x 
 - \frac{\zeta}{2}q_1^2
\end{equation}
which is, explicitly written,
\begin{eqnarray} \label{gbv2}
v_2 &=& \frac{\zeta}{2}
 + \frac{(1+i\sqrt{3})(4q_1^2+3\zeta^2)}{4\,3^{1/3}
 \left(-9\zeta^3 - 36\zeta q_1^2  + 2\sqrt{3}
 \sqrt{27\zeta^4q_1^2 + 72\zeta^2q_1^4 - 16q_1^6}\right)^{1/3}}
\nonumber \\
&& +\frac{(1-i\sqrt{3})\left(-9\zeta^3 - 36\zeta q_1^2 +2\sqrt{3}
 \sqrt{27\zeta^4q_1^2 + 72\zeta^2q_1^4 - 16q_1^6}
 \right)^{1/3}}{4\,3^{2/3}}
\end{eqnarray}
Next one can solve equations for $\tilde{M}^i$. Note that one of them
is not independent because of (\ref{cmm1}), but this can be easily
treated, e.g., by using Lagrange multiplier method. One obtains
\begin{equation}
0 = c_{ij}e^i\tilde{M}^j \left[ c_{nk}\tilde{M}^k
 \left(c_{ij}e^i\tilde{M}^j\right) - 2 c_{nk}e^k \right]
\end{equation}
from which follow conditions
\begin{equation} \label{gbmeq}
0 = c_{ij}e^i\tilde{M}^j \qquad \mbox{or} \qquad
(c_{nk}\tilde{M}^k)(c_{ij}e^i\tilde{M}^j) = 2 c_{nk}e^k 
\end{equation}
From the third equation in (\ref{gbeom}) (for $S$) we obtain
\begin{equation} \label{gbgee}
\tilde{G}_{ij}e^ie^j = v_1 +\frac{3}{2}\zeta\frac{v_1}{v_2}
\end{equation}
Last equation in (\ref{gbeom}), which defines electric charges, gives
\begin{equation} \label{gbqeom}
q_i = S\frac{v_2^{3/2}}{v_1}\tilde{G}_{ij}e^j
\end{equation}
which, together with (\ref{gijK3}) and (\ref{gbmeq}) gives
\begin{equation} \label{gbqce}
q_i = \mp S\frac{v_2^{3/2}}{v_1}c_{ij}e^j
\end{equation}
where the upper (lower) sign is when first (second) condition in 
(\ref{gbmeq}) applies. For the entropy we need $q_ie^i$, which from
(\ref{gbqeom}) is
\begin{equation} \label{gbqiei}
q_ie^i = \frac{v_2^{3/2}}{v_1}S\tilde{G}_{ij}e^ie^j
\end{equation}
We need a solution for the dilaton $S$ which is obtained by
contracting (\ref{gbqce}) with $q_kc^{ki}$. The result is
\begin{equation}
S = \frac{v_1}{v_2^{3/2}}\left|
 \frac{2q_ic^{ij}q_j}{\tilde{G}_{ij}e^ie^j} \right|^{1/2}
\end{equation}
Using this in (\ref{gbqiei}) we finally get for the black hole entropy
\begin{equation} \label{gbentK3}
S_{BH} = 2\pi q_ie^i = 4\pi \sqrt{\tilde{G}_{ij}e^ie^j} 
 \sqrt{\frac{1}{2}\left|q_ic^{ij}q_j\right|}
= 4\pi \sqrt{v_1 +\frac{3}{2}\zeta\frac{v_1}{v_2}}
 \sqrt{\frac{1}{2}\left|q_ic^{ij}q_j\right|}
\end{equation}
where $v_1$ and $v_2$ are functions of $q_1$ and $\zeta$ given in
(\ref{gbv1}) and (\ref{gbv2}). Observe that here entropy is nontrivial
function of charge $q_1$ (obtained by solving cubic equation),
contrary to the case of SUSY corrections which just introduce a
constant shifts.

For small black holes, i.e., when $q_1=0$, Eqs.\ (\ref{gbv1}), 
(\ref{gbv2}) and (\ref{gbgee}) simplify to
\begin{equation}
v_1 = \frac{v_2}{3} = \frac{\zeta}{2} \;,\qquad 
 \tilde{G}_{ij}e^ie^j = \zeta
\end{equation}
Plugging this in (\ref{gbentK3}) gives for the entropy
\begin{equation}
S_{BH} = 4\pi \sqrt{\frac{\zeta}{2}\left|q_ic^{ij}q_j\right|}
 \qquad \mbox{for } \; q_1=0
\end{equation}
which is exactly (\ref{eGBK3}).

\end{document}